\documentclass[aps,prb,article,floatfix,twocolumn,preprintnumbers,amsmath,amssymb,superscriptaddress]{revtex4}
\usepackage[usenames,dvipsnames]{color}

\usepackage{graphicx,xcolor}
\usepackage{dcolumn}
\usepackage{bm}
\usepackage{amsmath}    
\usepackage{graphics}
\usepackage[caption=false]{subfig}  
\usepackage{hyperref}   
\usepackage[permil]{overpic}
\usepackage{relsize}  
\usepackage[normalem]{ulem}
\usepackage{overpic}
\usepackage{xcolor}

\def\bfA{\ensuremath \mathbf A}

\def\bfE{\ensuremath \mathbf E}

\def\bfj{\mathbf j}
\def\bfB{\mathbf B}

\def\oh{\ensuremath \omega_{\mathrm H}}
\def\op{\ensuremath \omega_{\mathrm p}}
\def\ops{\ensuremath \omega_{\mathrm ps}}

\newcommand{\pderiv}[2]{\frac{\partial {#1}}{\partial {#2}}}

\newcommand{\mevh}      {\ensuremath{\mathrm{meV}/\hbar}}

\definecolor{green}{rgb}{0.3,0.5,0.}
\definecolor{blue}{rgb}{0,0.3,0.5}
\definecolor{orange}{rgb}{0.8,0.45,0.1}
\definecolor{blueAdd}{rgb}{0,0.2,0.8}

\begin{document}

\title{Higgs-mediated optical amplification in a non-equilibrium superconductor}

\author{Michele Buzzi}
\affiliation{Max Planck Institute for the Structure and Dynamics of Matter, 22761 Hamburg, Germany}

\author{Gregor Jotzu}
\affiliation{Max Planck Institute for the Structure and Dynamics of Matter, 22761 Hamburg, Germany}

\author{Andrea Cavalleri}
\affiliation{Max Planck Institute for the Structure and Dynamics of Matter, 22761 Hamburg, Germany}

\author{J. Ignacio Cirac}
\affiliation{Max Planck Institute for Quantum Optics, 85748 Garching, Germany}

\author{Eugene A. Demler}
\affiliation{Department of Physics, Harvard University, Cambridge, MA 02138, USA}

\author{Bertrand I. Halperin}
\affiliation{Department of Physics, Harvard University, Cambridge, MA 02138, USA}

\author{Mikhail D. Lukin}
\affiliation{Department of Physics, Harvard University, Cambridge, MA 02138, USA}

\author{Tao Shi}
\affiliation{CAS Key Laboratory of Theoretical Physics, Institute of Theoretical Physics,
Chinese Academy of Sciences, Beijing 100190, China}

\author{Yao Wang}
\affiliation{Department of Physics, Harvard University, Cambridge, MA 02138, USA}

\author{Daniel Podolsky\footnote{Corresponding author: podolsky@physics.technion.ac.il}}
\affiliation{Physics Department, Technion, 32000 Haifa, Israel}

\begin{abstract}
	The quest for new functionalities in quantum materials has recently been extended to non-equilibrium states, which are interesting both because they exhibit new physical phenomena and because of their potential for high-speed device applications. 
Notable advances have been made in the creation of metastable phases \cite{nasu2004photoinduced,rini2005photoinduced,morrison2014photoinduced,stojchevska2014ultrafast,
zhang2016cooperative,fausti2011light,mitrano2016possible,cantaluppi2018pressure,basov2017towards} and in Floquet engineering under external periodic driving\cite{Oka2019Floquet,McIver2018Light,Wang2013}. In the context of non-equilibrium superconductivity, examples have included the generation of transient superconductivity above the thermodynamic transition temperature \cite{fausti2011light,hu2014optically, mankowsyk2014nonlinear,mitrano2016possible,cantaluppi2018pressure}, the excitation of coherent Higgs mode oscillations \cite{matsunaga2013higgs,matsunaga2014light,matsunaga2017polarization,katsumi2018higgs} and the optical control of the interlayer phase in cuprates \cite{dienst2013optical,rajasekaran2016parametric}. Here, we propose theoretically a novel non-equilibrium phenomenon, through which a prompt quench from a metal to a transient superconducting state could induce large oscillations of the order parameter amplitude.  We argue that this oscillating mode could act as a source of parametric amplification of the incident radiation.  We report experimental results on optically driven K$_3$C$_{60}$ that are consistent with these predictions. The effect is found to disappear when the onset of the excitation becomes slower than the Higgs mode period, consistent with the theory proposed here. These results open new possibilities for the use of collective modes in many-body systems to induce non-linear optical effects.

\end{abstract}

\maketitle

\section{Introduction}

The Higgs mode is a fundamental collective excitation of systems with spontaneous symmetry breaking. It is a gapped excitation associated with oscillations of the amplitude of the order parameter. Examples of the Higgs mode in condensed matter are plentiful: it has been observed in the superconducting phase of NbSe$_2$ \cite{Sooryakumar:1980p809,LittlewoodVarma,VarmaHiggs} and of amorphous superconducting films \cite{sherman2015higgs}, in the dimerized antiferromagnet TlCuCl$_3$ \cite{ruegg}, in a variety of incommensurate charge density wave (CDW) systems \cite{ren,Pouget,yusupov}, and in cold bosonic gases near the superfluid to Mott insulator transition\cite{endres,Bissbort2011}.

\begin{figure}[h!]
    \includegraphics[width=\columnwidth]{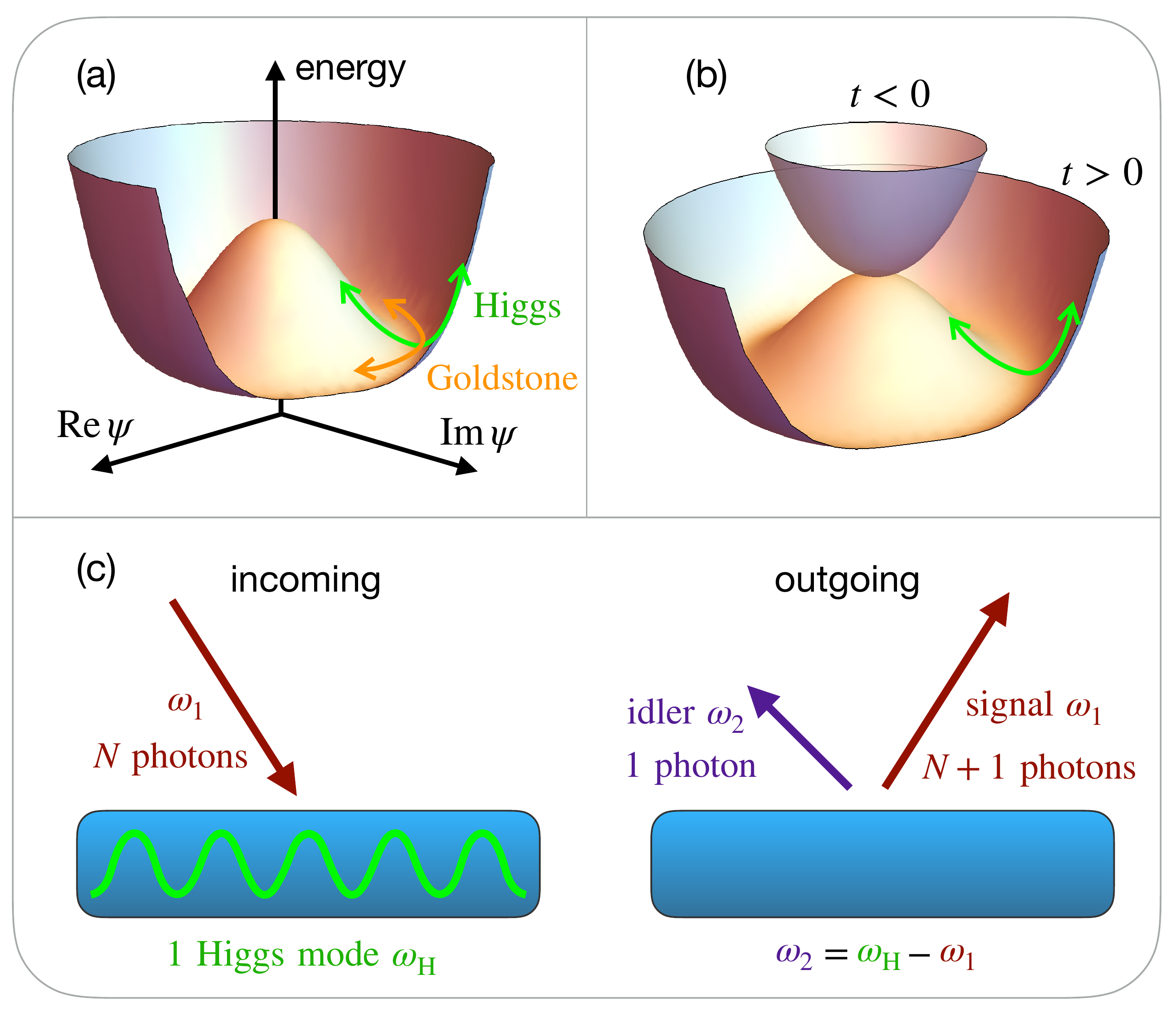}
    \caption{ (a) Schematic potential in a broken symmetry state. The collective excitations are Goldstone and Higgs modes.  
 (b)~A sudden change in system parameters at time $t=0$ leads to a rapid inversion in the potential, thus inducing Higgs oscillations. (c)~Light amplification in a superconductor with an excited Higgs mode. Left panel: A probe pulse containing $N$ photons of frequency $\omega_1$ is incident on a superconductor with an excited Higgs mode (represented by a wave).  Right panel: The output consists of $N+1$ photons of frequency $\omega_1$, plus a single photon of frequency $\omega_2$ traveling ``backward'', due to in-plane momentum conservation.  When many Higgs excitations undergo stimulated decay, a large number of both $\omega_1$ and $\omega_2$ photons are produced.
        \label{fig:fig_excited_higgs}} 
\end{figure}

In a number of experiments reported recently, superconductivity is created non-adiabatically after a rapid change in microscopic interactions, as induced by the application of a terahertz or mid-infrared (MIR) pump pulse \cite{denny2015proposed,kennes2017transient,nava2018cooling,babadi2017theory}. 
Although the dynamics of this process are not yet fully understood, we posit that the ``Mexican hat'' effective potential for the superconducting order parameter is established promptly after optical excitation, see Fig.~\ref{fig:fig_excited_higgs}(a). The appearance of large Higgs oscillations is then a natural attribute of such photo-induced superconductivity if the quench is fast compared to the frequency of the Higgs mode $\omega_H$, see Fig.~\ref{fig:fig_excited_higgs}(b) (the order parameter dynamics following a sudden quantum quench has been studied theoretically in a number of related contexts, see, {\em e.g.}, Refs. \onlinecite{BarankovLevitovSpivak,Yuzbashyan,BarankovLevitov,Calabrese2012}).

We argue that in this situation, the coherent collective mode can act as a source of parametric amplification\cite{clerk2010introduction} for oscillations of long-lived phase fluctuations resulting in an enhanced reflectivity for a time-delayed probe pulse.
This phenomenon can be qualitatively understood as a superconductor with an excited Higgs  mode, which then places a parametric modulation on low-frequency phase modes. We predict that the
reflected beam would then feature amplification of intensity at the original frequency $\omega_1$, as well as the generation of an idler signal at the complementary frequency $\omega_H-\omega_1$ (see Fig.~\ref{fig:fig_excited_higgs}(c)). We dub this phenomenon ``Higgs amplification''. 

We also report experimental results on optically driven K$_3$C$_{60}$, which support these predictions.  Using a pump-probe technique, we illuminate our samples with a pump pulse at 41 THz, and a probe pulse spanning frequencies between 1 to 7 THz.  Phase-resolved detection of the reflected signal allows us to reconstruct both the real and imaginary parts of the conductivity at the probe frequency. We find that when K$_3$C$_{60}$ is driven with pump pulses of suitable duration, the incident probe light is locally amplified near the surface for frequencies below 10 meV$/\hbar$. We analyze experimental results taking into account penetration depth mismatch between the pump and probe pulses.  We observe an anomalous enhancement of the reflectance, which in an homogeneously excited medium would result in 6\% amplification. For the
penetration depth of the probe beam of 700 nm this corresponds to the amplification coefficient $\alpha \sim 10^3$ cm$^{-1}$.

Our underlying theoretical considerations are presented in Sections \ref{sec:HiggsToAmp} and \ref{sec:nonlinearity}, below.  Experimental results are presented in Section \ref{sec:expResults}, and a comparison of theory and experiment is given in Section \ref{sec:comparison}.  Some implications of this work, and the outlook for further developments, are given in Section \ref{sec:outlook}.  Additional details of our analysis are given in three appendices.


\section{From Higgs oscillations to anomalous reflection}
\label{sec:HiggsToAmp}

The coupling between the Higgs mode and light can be understood by observing that in a general state with broken continuous symmetry, the Higgs mode can decay into a pair of Goldstone modes. 
For neutral superfluids, this process determines the lifetime of
Higgs excitations \cite{PAA}. In superconductors, the Goldstone mode describes phase fluctuations and therefore a charge current, which interacts directly with photons. Deep inside the material, these photons are gapped to the plasma frequency by the Anderson-Higgs mechanism, thus protecting the Higgs mode from decay.  However, near the surface, a Higgs excitation can decay into a pair of gapless vacuum photons whose frequencies add up to $\oh$. This corresponds to an effective term in the Hamiltonian
\begin{eqnarray}
{\cal H}_{\rm Higgs/photons} \sim \sum_{\omega_1+\omega_2=\oh}  (\, h\, a_{\omega_1}^\dagger a_{\omega_2}^\dagger \,+\, h^\dagger a_{\omega_1}a_{\omega_2} \,)
\label{Higgs_to_photons_schematic}
\end{eqnarray} 
where $h$ is a quantum field that annihilates a Higgs excitation and $a_\omega$ annihilates a vacuum photon of frequency $\omega$.
Hence, the energy of the excited Higgs mode is converted into entangled photon pairs. In the absence of the probe pulse, all photon pairs are generated equally. However, the incoming probe enhances the generation of pairs in which one of the photons matches the incident photons because of the bosonic stimulation factor. 

One can understand the origin of the coupling in Eq.~\eqref{Higgs_to_photons_schematic} from the following consideration. In superconductors, the diamagnetic coupling term between light and matter, 
$H_{\rm dia}=\frac{e^2\hbar^2}{2m c^2} n_{\rm s} {\bf A}^2,$ plays a key role, giving rise to the London equation for the current response to an electromagnetic field and to the Meissner effect. 
Here, $n_{\rm s}$ is the superfluid density which, for a coherently excited Higgs mode, is expected to oscillate at the Higgs frequency, $n_{\rm s}=n_{s,0}+\delta n \cos(\oh t)$.   In a quantized description of the electromagnetic field, the vector potential ${\bf A}$ is a linear combination of photon creation and annihilation operators, schematically of the form ${\bf A}\sim \sum_{\omega}(a_\omega+a_{\omega}^\dagger)$. This implies that the diamagnetic term gives rise to the term (\ref{Higgs_to_photons_schematic}), where $h$ is a quantum analog of the oscillating part of the superfluid density $\delta n$.
Interactions of this type are known to give rise to stimulated emission and parametric down conversion of light\cite{ou2008efficient}. 

These considerations give rise to the following mechanism of Higgs amplification. Consider an incident probe pulse composed of $N$ photons at frequency $\omega_1<\omega_H$ as it is reflected from a superconductor.  The term $a^\dagger_{\omega_1}a^\dagger_{\omega_2}$ creates a pair of photons, leading to a state with $N+1$ photons at $\omega_1$ and one  $\omega_2=\omega_H-\omega_1$ photon, see Fig.~\ref{fig:fig_excited_higgs}(c).  The amplitude of this process is enhanced by a Bose factor of $\sqrt{N+1}$.  As long as the Higgs mode remains excited, this process leads to amplification of $\omega_1$ photons and generation of $\omega_2$ photons, until the Higgs mode is depleted.  The net effect is outgoing light with two frequencies, $\omega_1$ and $\omega_2$, related by $\omega_1+\omega_2=\oh$, and the outgoing light at $\omega_1$ having a higher intensity than the incoming signal.

In typical superconductors, the Higgs frequency is smaller than the plasma frequency.  Then, incident light with $\omega_1<\omega_H$ does not penetrate deeply into the material, and Higgs amplification is a surface effect.  In particular, due to the evanescent nature of the waves inside the material, there are no phase matching conditions in this process.  This means that there isn't a discrete set of frequencies at which Higgs amplification is resonantly enhanced, and hence the level of amplification is expected to depend smoothly on the probe frequency $\omega_1$.  The frequency scale for the Higgs mode is comparable to the superconducting gap.  Hence, for many superconductors, Higgs amplification is expected to occur in the terahertz, a frequency range that is of great current interest for fundamental science and technology applications \cite{hwang2015review,graf2004terahertz}.  The next section presents a calculation of Higgs amplification based on a semiclassical description of photons using Maxwell's equations.
Our discussion is agnostic about the specific microscopic mechanism underlying light-induced superconductivity and, assuming that Higgs oscillations have been coherently excited, directly studies their effect on optical properties.

\section{Optical properties of a superconductor with an excited Higgs mode}
\label{sec:nonlinearity}

The optical properties of a superconductor with an excited Higgs mode are understood as follows. Consider Maxwell's equations combined with the relation between the electrical current in the material and electric field:
\begin{align}
{\mathbf \nabla}\times\bfB-\frac{\epsilon}{c^2}\pderiv{\,\bfE}{t}&=\mu_0\bfj\,,\label{eq:crossB}\\
{\mathbf \nabla}\times\bfE+\pderiv{\,\bfB}{t}&=0 \label{eq:crossE}\,.
\end{align}
Here $\epsilon$ is the dielectric constant arising from bound charges and ${\mathbf j}$ is the free current density.
For $z>0$, outside the superconductor, $\epsilon=\epsilon_{\rm out}$ (in vacuum $\epsilon_{\rm out}=1$, but we allow for interfaces with other media) and we assume the current ${\bf j}$ vanishes; for $z<0$,  $\epsilon=\epsilon_{\rm s}$, and the current satisfies the London equation,
\begin{align}
\bfj=\Lambda(t) \,{\bf v}_{\rm s}\,.\label{eq:London}
\end{align}
Here, ${\bf v}_{\rm s}=\frac{\hbar}{2e}\nabla\theta-\bfA$, where $\theta$ is the superconducting phase and $\bfA$ is the vector potential.   Note that $\partial_t {\bf v}_{\rm s}=\bfE$. In an equilibrium superconductor, $\Lambda(t)$ reduces to the static value $\Lambda_{\rm s}=\frac{e^2 n_{\rm s}}{m}$, where $n_{\rm s}$ is the  superfluid density.   More generally, in a time-modulated superconductor, Eq. (\ref{eq:London}) describes accurately the total current induced by a vector potential $\bf{A}$ in the limit where $\bf{A}$ and $\Lambda$ {\em vary slowly} compared to any microscopic frequencies such as the superconducting energy gap and the scattering rate $1/\tau$ for electrons in the normal state.  We use it here under the assumption that it is at least a good starting approximation for the frequencies of interest to us.

 In general, the value of $\Lambda$ will depend on the value of the superconducting energy gap. (Although the special case of an ideal clean superconductor at $T=0$ is an exception to this rule, we expect that the gap dependence will be manifest in the systems of interest to us, due to polaronic effects and coupling to impurities. \cite{MattisBardeen})  Since the Higgs mode represents a modulation in the energy gap, we expect it to induce a similar modulation in $\Lambda$.  Consequently, we may write
\begin{align}
\Lambda(t)=\Lambda_{\rm s}+\Lambda_{\rm m} e^{-i\oh t}+\Lambda_{\rm m}^* e^{i\oh t}\label{eq:Lambdat}
\end{align}
where $\Lambda_{\rm m}$ describes the amplitude of the modulation, and  $\oh$ is the Higgs mode frequency.    For a BCS superconductor, $\oh=2\Delta$, where $\Delta$ is the quasiparticle gap \cite{LittlewoodVarma}. 

Different frequencies mix due to the time dependence in $\Lambda$: incident light with frequency $\omega_1$ will produce outgoing light with frequencies $\omega_1$ and $\omega_2=\oh-\omega_1$, see Fig. \ref{fig:fig_excited_higgs}(c).   Mixing of the Higgs modulation and the signal beam is also expected to induce light at the frequency $\omega_3=\omega_1+\omega_H$.  We note, however, that the $\omega_3$ frequency lies well above the quasiparticle threshold $2\Delta$, and we expect that this channel of mode mixing will be suppressed relative to the channel at $\omega_2$.  This can be understood by observing that at such high frequencies the current should be carried primarily by quasiparticles, rather than supercurrents, and one might expect the quasiparticle current to be less sensitive than the supercurrent to modulations of the superconducting amplitude.  In the remainder of this paper, for the sake of simplicity, we assume that excitation of the $\omega_3$ mode is negligible, and we omit it entirely from our considerations.  Our results should be qualitatively applicable, however, as long as excitation of the $\omega_3$ mode is substantially weaker than that of the $\omega_2$ mode.


Prior to solving the reflection problem, it is useful to first consider the evanescent wave solutions inside the superconductor.   To simplify our discussion, we assume here that $\Lambda_m$ is spatially uniform inside the superconductor.  Non-uniformities in $\Lambda_m$ due to the short penetration depth of the exciting radiation will be taken into account in the processing of the experimental data, as detailed in Appendix \ref{sec:homogeneous}.  For the case of uniform $\Lambda_m$, the evanescent solutions are characterized by the spacetime dependence ($\bf{V}$ stands for $\bfj$, ${\bf v}_{\rm s}$, $\bfE$, or $\bfB$), 
\begin{align}
{\bf V} =\left({\bf V}_1 e^{-i\omega_1 t}+{\bf V}_2^* e^{i\omega_2 t} \right)\, e^{\kappa z}+{\mathrm{c.c.}} \,\label{eq:evanescent}
\end{align}
Note that the same spatial dependence appears for $\omega_1$ and $-\omega_2$ terms, since they mix with each other homogeneously in space.  This should be contrasted with the static case, where each frequency mode decays with its own $\kappa$.  Substituting Eq.~(\ref{eq:evanescent}) into Eq.~(\ref{eq:London}) and collecting terms with frequencies $\omega_1$ and $-\omega_2$ yields,
\begin{align}
 \bfj_1&=\Lambda_{\rm s}{\bf v}_{{\rm s},1}+\Lambda_{\rm m} {\bf v}_{{\rm s},2}^*=\Lambda_{\rm s}\frac{\bfE_1}{-i\omega_1}+\Lambda_{\rm m} \frac{\bfE_2^*}{i\omega_2},\label{eq:LondonMixed1}\\
  \bfj_2^*&=\Lambda_{\rm s}{\bf v}_{{\rm s},2}^*+\Lambda_{\rm m}^* {\bf v}_{{\rm s},1}=\Lambda_{\rm s}\frac{\bfE^*_2}{i\omega_2}+\Lambda_{\rm m}^* \frac{\bfE_1}{-i\omega_1}\,.\label{eq:LondonMixed2}
\end{align}
For linearly polarized light, $\bfE_\nu=E_{\nu}\hat{e}_x$, $\bfB_\nu=B_{\nu}\hat{e}_y$, and $\bfj_\nu=j_{\nu}\hat{e}_x$, where $\nu\in\{1,2\}$.  Then, Eq.~(\ref{eq:crossE}) yields $B_\nu=\frac{\kappa }{i\omega_\nu} E_\nu$, and Eq.~(\ref{eq:crossB}) becomes
\begin{align}
    \left(\begin{array}{c c} 
    \kappa^2 c^2+\omega_1^2\epsilon_1 & \Upsilon \omega_1/\omega_2\\
    \Upsilon^*\omega_2/\omega_1 & \kappa^2 c^2+\omega_2^2\epsilon_{\bar{2}}
    \end{array}\right)
    \left(\begin{array}{c c} 
    E_1 \\
    E_2^* 
    \end{array}\right)=0\,.
\end{align}
Here, $\Upsilon\equiv\Lambda_{\rm m}/\varepsilon_0$, $\epsilon_1\equiv\epsilon(\omega_1)$, and $\epsilon_{\bar{2}}\equiv\epsilon(-\omega_2)$, where $\epsilon(\omega)=\epsilon_{\rm s}-\Lambda_{\rm s}/(\varepsilon_0\omega(\omega+i 0^+)$ is the dielectric function of the superconductor [Note that, by changing the form of $\epsilon(\omega)$, one can generalize the discussion, {\em e.g.}, to introduce dissipation].  The allowed values of $\kappa$ are obtained by requiring the determinant of the above matrix to vanish,
\begin{align}
\kappa_\pm^2 c^2 =-\frac{\omega_1^2\epsilon_1+\omega_2^2\epsilon_{\bar{2}}}{2}\pm \sqrt{\frac{\left(\omega_1^2\epsilon_1-\omega_2^2\epsilon_{\bar{2}}\right)^2}{4}+|\Upsilon|^2},\label{eq:kappapm}
\end{align}
and the fields inside the superconductor satisfy,
\begin{align}
\frac{E_{1,\pm}^{\rm t}}{B_{1,\pm}^{\rm t}}&=\frac{i\omega_1}{\kappa_\pm}\equiv \eta_\pm\label{eq:etapm}\\
\frac{E_{2,\pm}^{{\rm t}*}}{B_{1,\pm}^{\rm t}}&=\frac{-i\omega_2\left(\kappa_\pm^2 c^2+\omega_1^2\epsilon_1 \right)}{\Upsilon \kappa_\pm}\equiv \phi_\pm\label{eq:phipm}\\
\frac{B_{2,\pm}^{{\rm t}*}}{B_{1,\pm}^{\rm t}}&=\frac{\left(\kappa_\pm^2 c^2+\omega_1^2\epsilon_1\right)}{\Upsilon}\equiv \gamma_\pm\label{eq:gammapm}
\end{align}
where the superscript ${\rm t}$ indicates that these act as transmitted fields in the reflection problem.

Next, we consider a normally-incident signal beam of frequency $\omega_1$, and reflected beams at frequencies $\omega_1$ and $\omega_2$, with 
\begin{align}
\bfE^{\rm i}_1,\bfB^{\rm i}_1&\propto e^{-i\omega_1(t+z\sqrt{\epsilon_{\rm out}}/c)}\,, \label{eq:incident1}\\
\bfE^{\rm r}_\nu,\bfB^{\rm r}_\nu&\propto e^{-i\omega_\nu(t-z\sqrt{\epsilon_{\rm out}}/c)} \label{eq:reflected1}\,.
\end{align}
where $\nu\in\{1,2\}$.  The fields are linearly polarized as before,  $\bfE_\nu=E_{\nu}\hat{e}_x$, $\bfB_\nu=B_{\nu}\hat{e}_y$, and Maxwell's equations constrain them to satisfy $E_1^{\rm i}=-(c/\sqrt{\epsilon_{\rm out}})B_1^{\rm i}$ and $E_\nu^{\rm r}=(c/\sqrt{\epsilon_{\rm out}})B_\nu^{\rm r}$.
Within the superconductor, the transmitted fields are superpositions of two evanescent waves of the form (\ref{eq:evanescent}) with $\kappa=\kappa_\pm$, whose amplitudes are fixed by boundary conditions imposed independently on each frequency:
\begin{align}
B_1^{\rm i}+B_1^{\rm r}&=B_{1,+}^{\rm t}+B_{1,-}^{\rm t}\label{eq:B1boundary}\\
E_{1}^{\rm i}+E_{1}^{\rm r}&=E_{1,+}^{\rm t}+E_{1,-}^{\rm t}\label{eq:E1xboundary}\\
B_2^{{\rm r}*}&=B_{2,+}^{{\rm t}*}+B_{2,-}^{{\rm t}*}\label{eq:B2boundary}\\
E_{2}^{{\rm r}*}&=E_{2,+}^{{\rm t}*}+E_{2,-}^{{\rm t}*}\label{eq:E2xboundary}
\end{align}
Combining these with Eqs.~(\ref{eq:etapm}), (\ref{eq:phipm}), and (\ref{eq:gammapm}) yields
\begin{align}
r&\equiv\frac{B_1^{\rm r}}{B_1^{\rm i}}=\frac{(1+\zeta)+\sqrt{\epsilon_{\rm out}}(\eta_++\zeta\eta_-)}{(1+\zeta)-\sqrt{\epsilon_{\rm out}}(\eta_++\zeta\eta_-)}\,,\label{eq:R11}\\
r_{12}&\equiv\frac{B_2^{{\rm r}*}}{B_1^{\rm i}}=\frac{2 (\gamma_++\zeta\gamma_-)}{(1+\zeta)-\sqrt{\epsilon_{\rm out}} (\eta_++\zeta\eta_-)}\,,\label{eq:R12}
\end{align}
where
\begin{align}
\zeta=-\frac{\gamma_+-\sqrt{\epsilon_{\rm out}}\phi_+}{\gamma_--\sqrt{\epsilon_{\rm out}}\phi_-}\,.\label{eq:zeta}
\end{align}
Eqs.~\eqref{eq:R11} and \eqref{eq:R12} are the main theoretical results of this paper:  $r$ is the reflection amplitude of the signal beam, whereas $r_{12}$ is the amplitude of the emitted idler mode at the down-converted frequency $\omega_2=\oh-\omega_1$, normalized by the amplitude of the incident signal beam.

\begin{figure}
    \subfloat[\label{fig:R1_normal}]{\begin{centering}
        \hspace{2 pt}    
	\begin{overpic}[width = 0.4\textwidth]{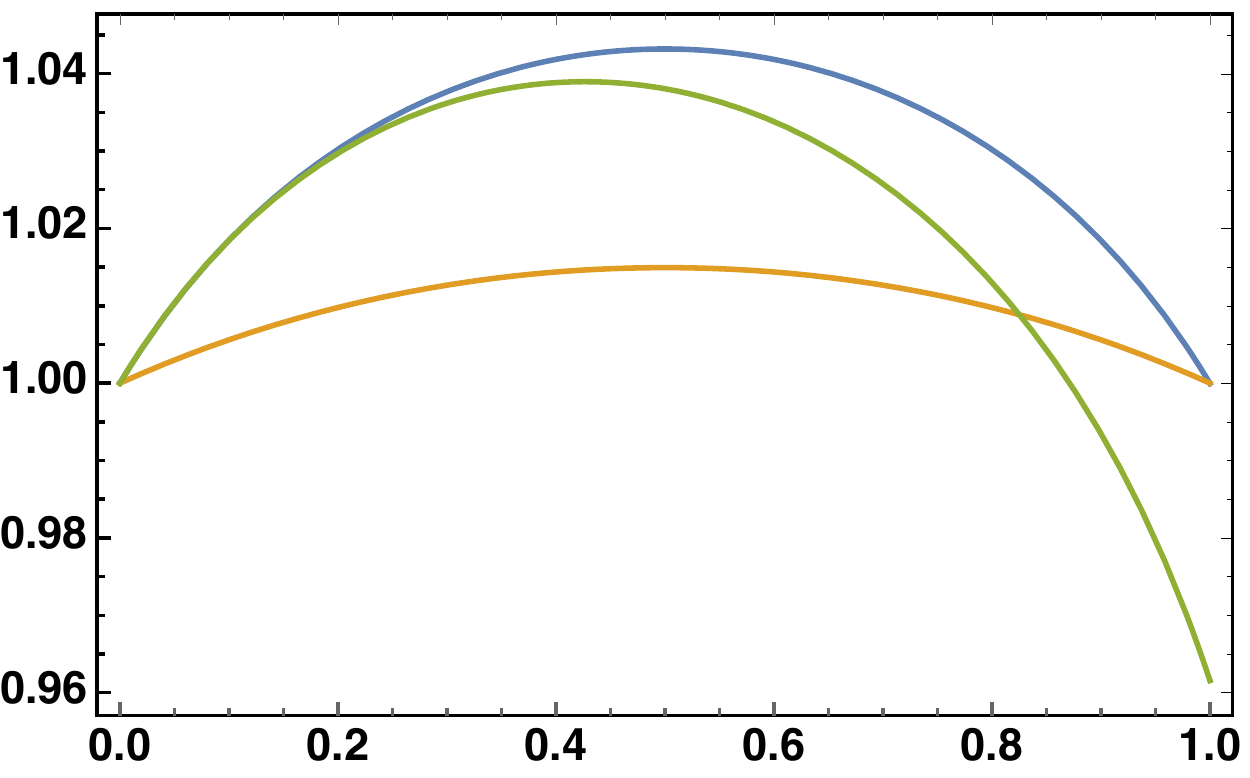}
	      \put(-105,550){\large (a)}
                \put(-85,300){\large \rotatebox{90}{$R$}}
                \put(490,-40){$\omega_1/\oh$}
		\put(110,550){\color{blue} $\oh=0.8\ops$}
		\put(260,460){\color{green} $\oh=0.8\ops$, $\sigma_r\ne 0$}
		\put(340,340){\color{orange} $\oh=0.5\ops$}
            \end{overpic}
    \end{centering}}
    \hfill{} 
    \subfloat[\label{fig:R2_normal}]{\begin{centering}
	\hspace{2 pt}
            \begin{overpic}[width = 0.4\textwidth]{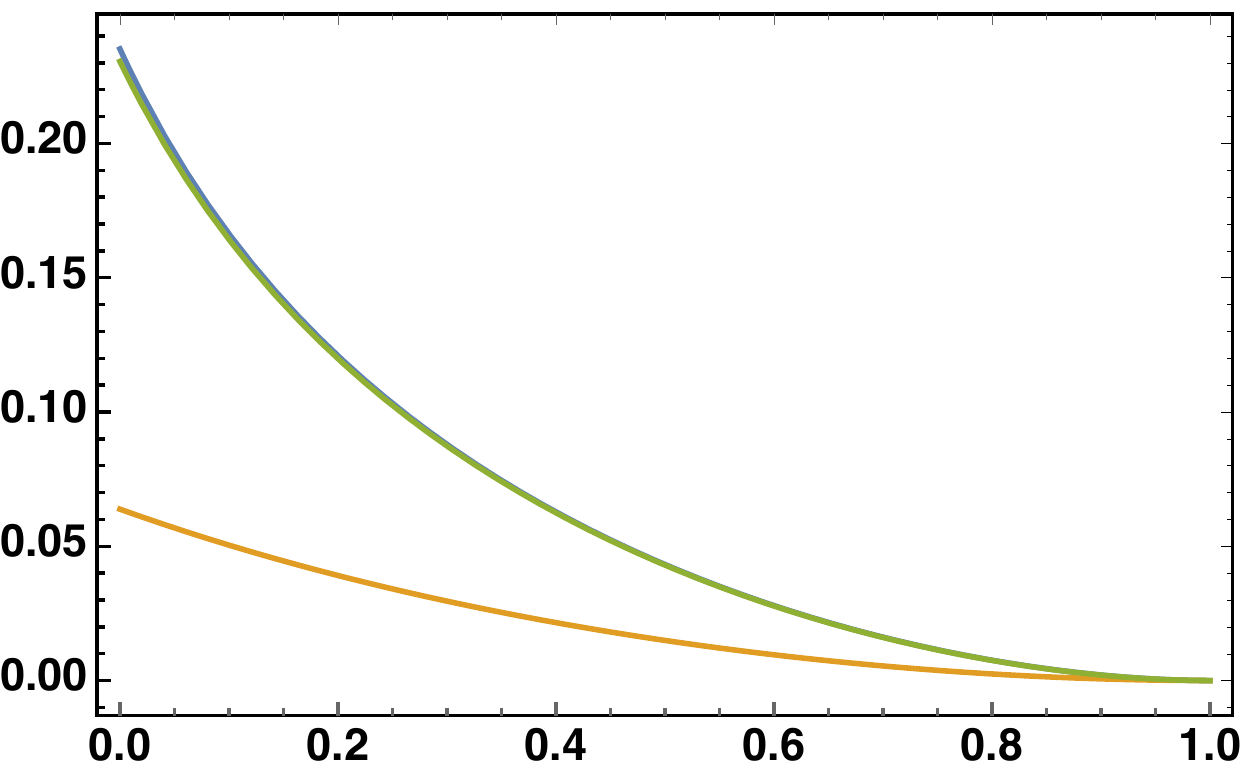}
                \put(-105,550){\large (b)}
	      \put(-85,250){\large \rotatebox{90}{$R_{12}$}}
                \put(490,-40){$\omega_1/\oh$}
		\put(180,460){\color{blue} $\oh=0.8\ops$}
		\put(260,360){\color{green} $\oh=0.8\ops$, $\sigma_r\ne 0$}
		\put(130,220){\color{orange} $\oh=0.5\ops$}
            \end{overpic}
            \par\end{centering}}
    \caption{(a) Reflection coefficient $R$ and (b) downconversion intensity $R_{12}$ as a function of the signal frequency $\omega_1$. We  assume $\epsilon_{\rm s}=5$, $\epsilon_{\rm out}=5.62$ (corresponding to diamond), and $\Lambda_{\rm m}=0.4\Lambda_{\rm s}$.  Two different values of the Higgs frequency were taken, $\oh=0.5\ops$ and $\oh=0.8 \ops$, where $\ops=\sqrt{\Lambda_{\rm s}/(\varepsilon_0\epsilon_{\rm s})}$ is the superconducting plasma frequency.  The green curves, corresponding to $\oh=0.8 \ops$, illustrate the effects of dissipation, taken into account by 
assuming $\epsilon(\omega)=\epsilon_{\rm s}-\sigma(\omega)/(i\varepsilon_0\omega)$ with $\sigma(\omega)=\sigma_{\rm r}+i\Lambda_{\rm s}/(\omega+i0^+)$, where $\sigma_{\rm r}$ is a real constant; the blue and orange curves include no dissipation, $\sigma_{\rm r}=0$. Results are for the model in which $\Lambda_m$ is uniform in space.
    		\label{fig:R_normal} }
\end{figure}
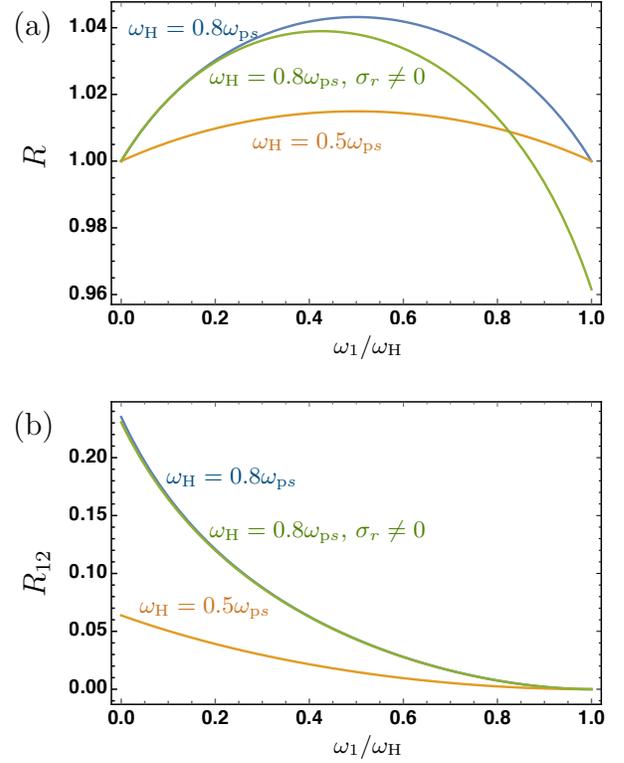

Figure \ref{fig:R_normal} shows $R=|r|^2$ and $R_{12}=|r_{12}|^2$ as a function of $\omega_1$, in the case $\oh< \ops$, where $\ops=\sqrt{\Lambda_{\rm s}/(\varepsilon_0\epsilon_{\rm s})}$ is the superconducting plasma frequency of the material.  Note that, in the absence of dissipation,  there is amplification $R>1$  over the entire range $0<\omega_1<\oh$, and the maximum amplification occurs at $\omega_1=\oh/2$.   One can study the effect of dissipation by writing $\epsilon(\omega)=\epsilon_{\rm s}-\sigma(\omega)/(i\varepsilon_0 \omega)$ in Eqs.~(\ref{eq:R11}) and (\ref{eq:R12}) and adding a real part to the conductivity function $\sigma(\omega)$.  This suppresses the reflectivity and also reduces the frequency $\omega_1$ at which the maximum occurs, as shown in Fig.~\ref{fig:R_normal}(a).

Note that the net amplification in Fig. \ref{fig:R_normal} is small, of the order of a few percent.  In order to understand this, focus on $\omega_1=\omega_2=\oh/2$, where the maximum is obtained, with value
\begin{align}
R_{\mathrm{max}}=1+\frac{\epsilon_{\rm out}\,\oh^2\left(D_+-D_-\right)^2}{\left[\epsilon_{\rm out}\,\oh^2+D_+D_-\right]^2}\,,
\end{align}
where
$D_\pm=\sqrt{4 \left(\Lambda_{\rm s}\pm |\Lambda_{\rm m}|\right)/\varepsilon_0-\epsilon_{\rm s}\,\oh^2}\,.$
For $\oh^2\ll \Lambda_{\rm s}$ and $\Lambda_{\rm m}\ll \Lambda_{\rm s}$, this can be expanded to obtain 
\begin{align}
R_{\mathrm{max}}=1+\frac{|\Lambda_{\rm m}|^2}{\Lambda_{\rm s}^2}\frac{\epsilon_{\rm out}\,\oh^2}{4\,\epsilon_{\rm s}\ops^2}
\,.\label{eq:maxR}
\end{align}
The factor $|\Lambda_{\rm m}|^2/\Lambda_{\rm s}^2$ expresses the fact that the amplification is proportional to the intensity of the modulation.  The factor $\epsilon_{\rm out}\oh^2/(4\epsilon_{\rm s}\ops^2)$ is the square of the London penetration depth divided by the wavelength of the incident light; it expresses the fact that amplification is weak if the light cannot probe deeply into the superconductor.  This suggests that, in order to enhance the amplification, one could consider instances in which the light can probe a larger region of the superconductor prior to being reflected.  Two possible approaches to achieve this come to mind: First, by using incoming light with a shallow incidence angle to the sample, thus introducing geometrical factors which enhance the effect; second, by studying systems in which the Higgs frequency exceeds the plasma frequency.  These possibilities will be discussed elsewhere \cite{PodolskyuUnpubHiggsAmp}.  Note that, in a realistic superconductor, the penetration of light is controlled by the total plasma frequency $\op$, which receives contributions from all charge carriers, not just the superconducting ones.  It is this total plasma frequency that presumably controls the strength of the Higgs amplification in Eq. (\ref{eq:maxR}).

In the next section, we will present experimental evidence for Higgs amplification in the superconductor K$_3$C$_{60}$.  This is a natural candidate system for Higgs amplification.  In this material,  the low density of electrons (three per C$_{60}$ molecule) and the weak hopping between C$_{60}$ molecules conspire to yield an anomalously small plasma frequency $\op=72 \mevh$.  Optical excitation at mid-infrared wavelengths has been shown to transform the high-temperature $(T\gg T_c)$ metallic phase of K$_3$C$_{60}$ into a transient non-equilibrium state with the same optical properties as the low temperature superconductor ($T<T_c$). The transient state, which is thought to be a photoinduced non-equilibrium superconductor, displays a saturated reflectivity ($R=1$), a gap in the real part of the optical conductivity $\sigma_1$ and a divergent low-frequency imaginary conductivity $\sigma_2$ \cite{mitrano2016possible}.
In order to estimate the zero-temperature gap, $2\Delta$, at the light-induced state, we use the onset temperature for light-induced superconductivity, $T_{\rm light-induced}\approx 100$ K, which yields $2\Delta\approx 30$ meV \cite{mitrano2016possible}.  This gives an estimate for $\oh$, which is expected to be reduced from this value at non-zero temperature (in the fit to our theory, we find $\oh$ is of the order of $24$ \mevh, see Sec.~\ref{sec:comparison}).  Hence, K$_3$C$_{60}$ combines a relatively large value of $\oh/\op$, as necessary to enhance the reflectance in Eq.~(\ref{eq:maxR}), together with the possibility for rapid quenches into the superconducting state by pump pulses, as needed to induce large Higgs oscillations.  In particular, by modifying the type of quench protocol used, one may control the amplitude of Higgs oscillations.

The Higgs mode should be rather insensitive to the lack of long-range phase coherence, which is expected not to exceed a few coherence lengths in the photoinduced superconducting state.   Hence, we expect the mechanism of Higgs amplification to occur in this case, despite the lack of global phase coherence.

\section{Experimental results}
\label{sec:expResults}

We follow the same protocol described in Ref. \onlinecite{mitrano2016possible, cantaluppi2018pressure}, to photo-induce the transient optical properties of a non-equilibrium superconductor in K$_3$C$_{60}$ but using shorter and more intense pulses. To test the hypotheses discussed in the theory section above, we additionally analyzed conditions in which the quench was made slower with respect to the Higgs mode frequency of the photo-induced state. The pump pulse FWHM duration $\tau$ was tuned to different values between 100 fs and 1.8 ps. This range of pulse durations is interesting because it crosses a characteristic time scale $\tau^*=h/(24\,\mathrm{meV})\sim 172$ fs, which corresponds to the period of the amplitude (Higgs) mode in the photo-induced superconductor. 
Note that in our experimental geometry the idler mode is not detected. In future studies, this mode could provide a measurement of the Higgs frequency and would allow for distinguishing Higgs amplification from other types of parametric amplification \cite{cartella2018phonons,liu2013phasons}.

In the excitation regime explored in this experiment, when the pump pulse is significantly longer than $\tau^*$, the transient state displays a reflection coefficient at the surface that is saturated at $R=1$ for all frequencies below $\sim $10meV. By reconstructing the complex optical conductivity with the same procedure used previously\cite{cantaluppi2018pressure}, we extracted a gapped real optical conductivity $\sigma_1$ at all frequencies $\omega<10$ meV and a divergent low-frequency imaginary conductivity $\sigma_2$, indicative of a light-induced superconducting state. Our experiment reveals that as the pulse duration is made progressively shorter than $\tau^*$, the transient state acquires a reflection coefficient that is larger than $R=1$ immediately after the pump, indicative of optical amplification through the non-adiabatic creation of a superconducting state. The real part of the optical conductivity $\sigma_1$ is negative at all frequencies $\omega<10$ meV while its imaginary part $\sigma_2$ remains divergent. 
Conceptually, the observed dependence on the pulse duration can be understood from the following consideration. Although the underlying mechanism of photo-induced superconductivity is still the subject of debate, we assume as a first approximation that the ``effective'' final state Hamiltonian experienced by the low-energy electrons depends only on the total pulse energy, and not its duration  (later, we will show evidence that the shorter pulses actually do drive the superconductivity more strongly and induce a slightly larger superfluid density, although this is a weak effect).  We furthermore assume that the superconducting state lasts longer than the probe sequence. Hence qualitatively we  assume that by controlling the pump pulse duration we preserve the final effective Hamiltonian for electrons but change the rate at which the microscopic parameters are modified.  In the case of a superconductor we expect that the Higgs-amplitude mode gets strongly excited when the interaction strength is modified on a time scale shorter than $\tau^* = h/2\Delta$.


\begin{figure}[]
    \includegraphics[width=\columnwidth]{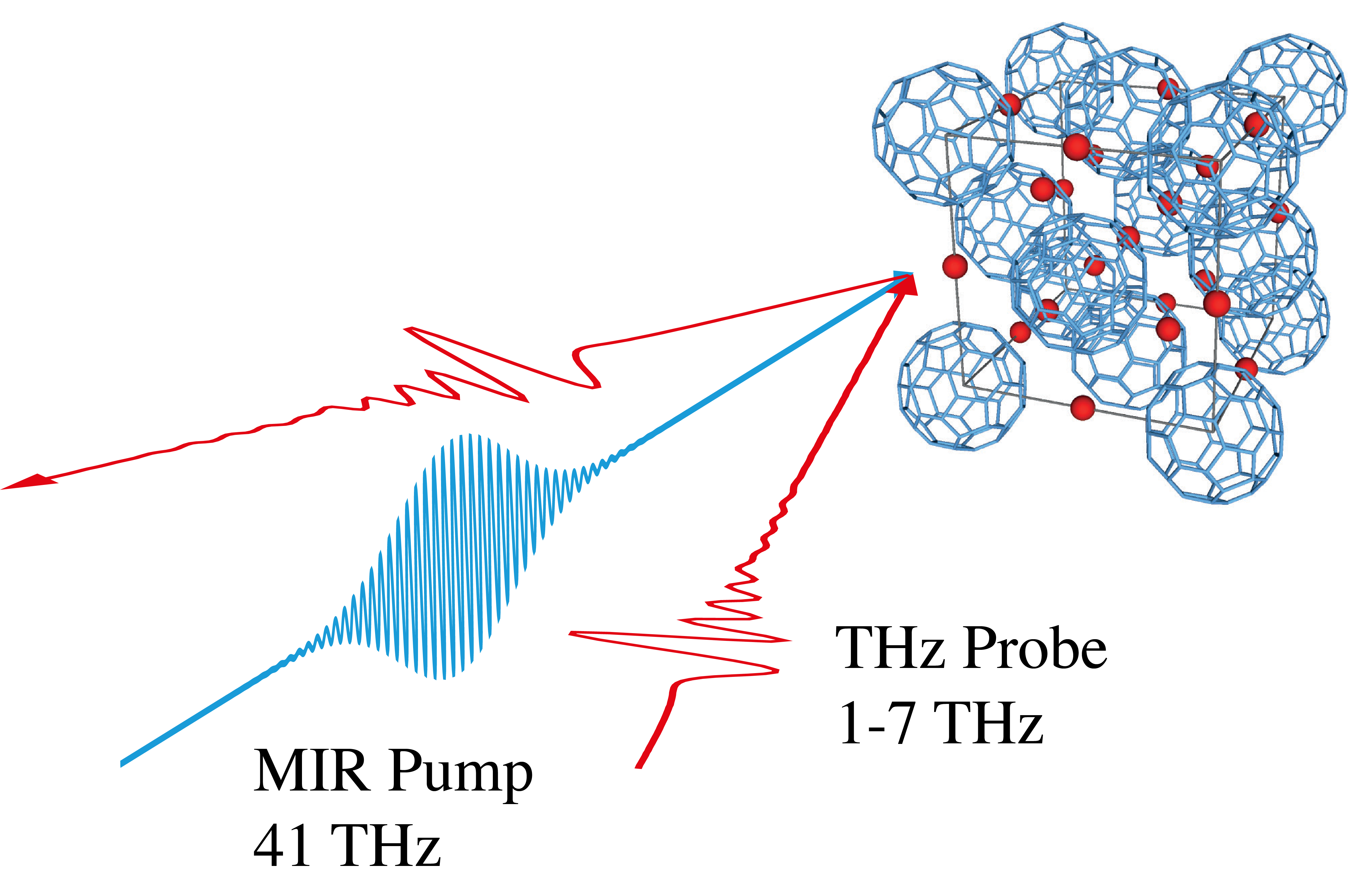}
    \caption{Sketch of the experimental geometry. K$_3$C$_{60}$ is excited with vertically polarized mid-infrared pulses. The THz probe pulses are polarized in the horizontal plane.
        \label{fig:ExpFig1}} 
\end{figure}

K$_3$C$_{60}$ polycrystalline powders were excited at normal incidence with 170 meV linearly-polarized mid-infrared pulses. Their duration was tuned from 100 fs to 1.8 ps by chirping them through linear propagation in transparent and highly dispersive CaF$_2$ rods. For all pulse durations, the pulse energy and the number of incident photons were maintained constant. The transient low-frequency optical properties of photo-excited K$_3$C$_{60}$ were retrieved as a function of pump-probe delay using transient terahertz time-domain spectroscopy using THz pulses with a bandwidth that ranged from 1 to 7 THz. These probe pulses were made to strike the sample at near normal incidence, with a $7^\circ$ incidence angle (see Figure \ref{fig:ExpFig1}), and they were $p$-polarized, that is, with the electric field perpendicular to that of the MIR pump pulses. The measurement of the electric field reflected from the sample yielded a phase-resolved measurement of the reflection coefficient and through it the complex optical properties. The penetration depth of the mid-infrared pump (200nm) was shorter than that of the THz probe (600-900nm). To account for this, the data was analyzed as discussed in Appendix \ref{sec:homogeneous} in order to obtain the reflectivity corresponding to an effective semi-infinite and homogeneously pumped medium.

\begin{figure*}[]
    \includegraphics[width=0.9\textwidth]{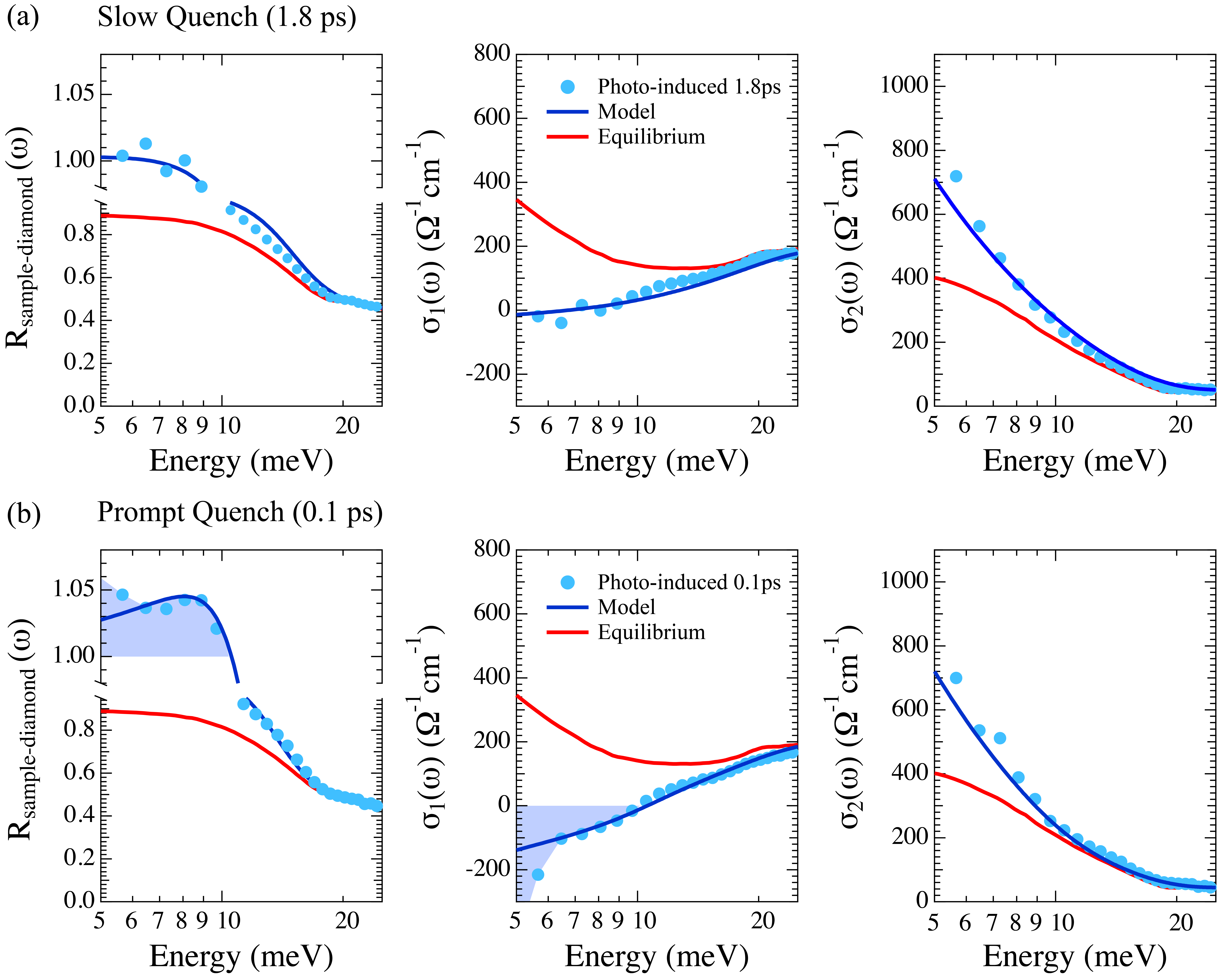}
    \caption{ (a)  Reflectivity and complex optical conductivity (sample-diamond interface) of K$_3$C$_{60}$ measured at equilibrium (red solid curves) and at the peak of the pump-probe response (light-blue dots) with pump pulse duration of 1800fs. For non-equilibrium systems we present inferred local quantities at the sample-diamond interface as discussed in Section V and Appendix A. (b) Same quantities, measured with pump pulse duration of 100fs. The shaded area highlights the frequency range where amplification is observed. All data were taken at $T$=100K and at a fluence of 4.5mJ/cm$^2$. The blue solid curves are the the optical conductivity and reflectivity calculated from the theoretical model taking into account Higgs modulations. The corresponding fit parameters are shown in Table \ref{tab:modelParameters}.
        \label{fig:ExpFig2}} 
\end{figure*}

Figures \ref{fig:ExpFig2}(a) and \ref{fig:ExpFig2}(b) compare the transient optical properties of K$_3$C$_{60}$ upon photoexcitation with intense mid-infrared pulses of 1.8 ps and 100fs duration, respectively. The red curves report the optical properties of the equilibrium metallic state, whilst the light-blue dots represent those of the transient state, measured at the peak of the response. For longer excitation pulses [1.8 ps, Fig.~\ref{fig:ExpFig2}(a)] the transient optical properties resemble those of the equilibrium superconductor with a reflectivity saturated at $R=1$, a gapped real conductivity ($\sigma_1 \approx 0$) and a divergent imaginary conductivity for all frequencies below $\sim $10meV. For $\tau=100$fs [Fig.~\ref{fig:ExpFig2}(b)] we find that below $\sim$10 meV, the reflectivity becomes larger than $R=1$, reaching an average value in the gapped region of $\sim 1.04$, with a maximum of $\sim 1.06$.  The extracted real part of the optical conductivity is correspondingly negative, indicative of negative dissipation. These two observations suggest amplification of the incoming low frequency THz probe light. Importantly, the imaginary part of the optical conductivity maintained a $1/\omega$ behavior below $\sim 10$meV, indicating the superconducting nature of the transient state. 

\begin{figure}[]
\begin{overpic}[width=0.85\columnwidth]{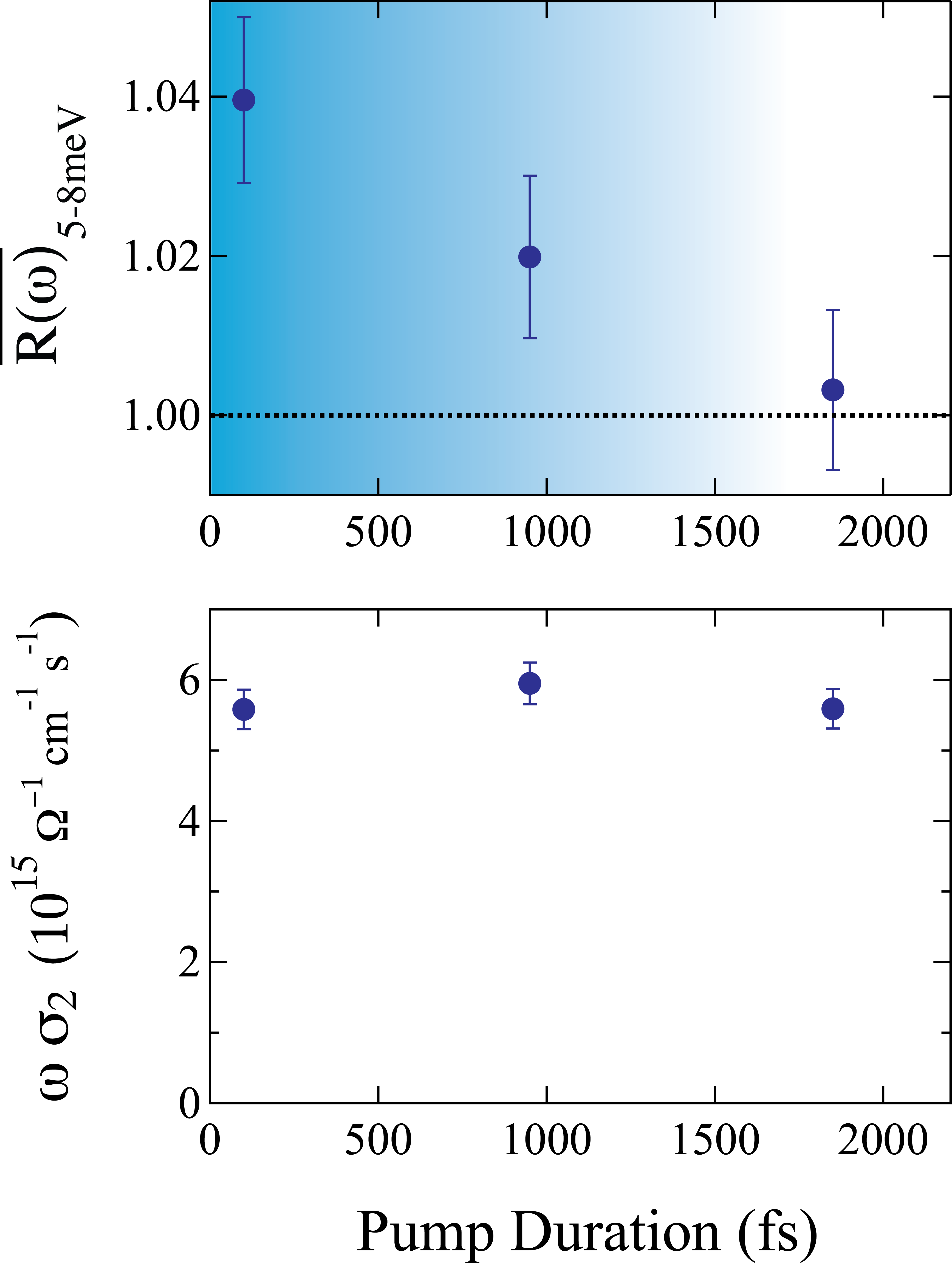}
                \put(190,625){\large (a)}
               \put(190,150){\large (b)}
	\end{overpic}
    \caption{(a) Inferred local transient reflectivity (sample-diamond interface), averaged between 5-8meV, as a function the pump pulse duration. The blue-shaded region indicates the range of pulse durations where amplification is observed. (b) $\omega\sigma_2(\omega\to 0)=\Lambda_{\rm s}$, which is proportional to the transient superfluid density, extracted by a low frequency $1/\omega$ fit to $\sigma_2(\omega)$. All data were taken at $T$=100K and a constant pump fluence of 4.5mJ/cm$^2$.
        \label{fig:ExpFig3}} 
\end{figure}

The evolution of the optical properties for pump pulses either shorter or longer than $\tau^*$ can be captured by the average value of the reflectivity in the 5-8meV-frequency range and the superfluid density extracted by a $1/\omega$ fit to the imaginary part of the optical conductivity at low frequencies.  Figure  \ref{fig:ExpFig3}(a) and  \ref{fig:ExpFig3}(b) show the evolution of these two quantities as a function of the duration of the excitation pulse (Fig.~ \ref{fig:ExpFig3}(b)).  The average reflectivity in the gapped region decreases from $\sim 1.04$ to $\sim 1$, as the pump pulse duration varies from 100 fs to 1.8 ps.  The blue shaded area in the top panel highlights the regime where light amplification is observed.  At the same time, the superfluid density of the photoinduced superconductor does not appear to strongly depend on the pulse duration of the excitation pulse.

\section{Comparison of theory and experiment}
\label{sec:comparison}

The experimental data does not give a direct measurement of the Higgs frequency  -- in the light-induced superconducting state, the gap measured in $\sigma_1(\omega)$ coincides with the lower edge of the mid-infrared absorption peak of K$_3$C$_{60}$, suggesting that the superconducting gap $2\Delta$ is hidden under the spectral weight of this broad peak.  However, there are two independent observations that give consistent estimates for $\oh$.  First, as argued earlier, the maximum of light amplification occurs slightly below $\oh/2$. The reflectivity measured with the shortest excitation pulse ($\tau=100$ fs) shown in Fig.~\ref{fig:ExpFig2}(a) would then suggest that $\oh$ is somewhat bigger than $20 \mevh$. Second, as mentioned earlier, the onset temperature of the light-induced superconductivity corresponds to a gap $2\Delta(T=0)$ of $30\mevh$ for the zero-temperature transient superconductor.  This gives an upper bound on $\oh$, which tracks $2\Delta(T)$.

The measured optical conductivity $\sigma(\omega)$ has Higgs modulations built in.  In order to make a comparison between theory and experiment, we need to model the optical conductivity of the  static superconducting state, $\tilde{\sigma}$.  We parametrize $\tilde{\sigma}$ as a sum of Drude and Lorentzian peaks:
\begin{eqnarray}
\tilde{\sigma}(\omega)=\frac{\Lambda_{\rm s}}{\gamma_{\rm D}-i \omega} + \sum_{n=1}^3  \frac{B_n\, \omega}{i(\Omega_n^2-\omega^2)+ \gamma_n\, \omega} \label{eq:Lorentzian}
\end{eqnarray}
The Lorentzians represent the broad mid-infrared absorption peak in K$_{3}$C$_{60}$ \cite{RiceChoi}.  For simplicity, we assume this peak to be unaffected by the onset of superconductivity, and fix the parameters $B_n$, $\Omega_n$, and $\gamma_n$ by fitting to the conductivity of the equilibrium normal state, as discussed in App.~\ref{sec:sigmaEq}.  For the light-induced superconducting phase, we maintain the same values for the Lorentzian peaks, while for the Drude peak we replace $\gamma_{\rm D}\to 0^+$ and allow $\Lambda_{\rm s}$, which plays the role of the static superfluid density, to vary.   

Given the static conductivity $\tilde{\sigma}$, we use the results of Section \ref{sec:nonlinearity} to model the Higgs amplification phenomenon. We apply  Eq.~(\ref{eq:R11}) to compute the complex reflection coefficient $r$ at the signal frequency,  in which we take $\epsilon(\omega)=\epsilon_{\rm s}-\tilde{\sigma}(\omega)/(i\varepsilon_0\omega)$.  By inverting the Fresnel equation, $r$ is then expressed as an effective optical conductivity $\sigma(\omega)$, see Sec.~\ref{sec:RtoSig}, allowing us to make a comparison of the full complex response of the system.  

We find that it is possible to describe the experimental data well by taking $\oh=24$ \mevh.  Then, for each value of $\tau$, there are only two parameters we allow to change, the superfluid density $\Lambda_{\rm s}$ and the amplitude of Higgs modulations, $\Lambda_{\rm m}/\Lambda_{\rm s}$. We emphasize that all other parameters in the model are determined in an unbiased manner by comparison with the equilibrium optical properties of the normal state. The blue solid curves in figure \ref{fig:ExpFig2}(a,b) shows the complex optical conductivity and the reflectivity of the superconductor with parameters as chosen in Table \ref{tab:modelParameters}. These results show good agreement with measurements, despite the limited number of fitting parameters used and the simplicity of the model, which does not take into account the full microscopic details of the system. In our fit,  we find that the Higgs modulation amplitude increases with decreasing $\tau$, in agreement with the expectation that shorter pulses give rise to more rapid quenches into the superconducting state, and therefore to larger Higgs oscillations.  In addition, we find that $\Lambda_{\rm s}$ is slightly larger than its equilibrium value in the normal state $\Lambda_{\rm s,eq}$.   This effect is larger for the shorter pulses. Even though the total pump pulse energy is maintained fixed, the shorter pulses have higher peak intensities, and can drive the superconductor more strongly since the pump drives the system nonlinearly.  Note that, in order to preserve sum rules, this requires spectral weight to transfer from higher energies and may be an indication that the plasmonic peak must be modified in a complete microscopic description of the phenomenon.

\begin{table}[h!]
  \begin{center}
    \begin{tabular}{c || c | c} 
      $\tau$ (fs) & $\Lambda_{\rm s}/\Lambda_{\rm s,eq}$ &  $\Lambda_{\rm m}/\Lambda_{\rm s}$ \\ \hline
      1800& 1.05 &  0.2 \\
      1000&1.11 &  0.37 \\
      100 & 1.17 & 0.47 \\
    \end{tabular}
    \caption{Model parameters for different pump pulse widths.}
    \label{tab:modelParameters}
  \end{center}
\end{table}

The goal of our theoretical analysis was to provide the simplest physical picture of Higgs amplification.  Therefore, we limited our discussion to the simplest case, in which the Higgs modulations were assumed to be monochromatic and spatially uniform.  In practice, one may expect broadening of the Higgs excitation in frequency, due to its finite lifetime, and in momentum, due to the non-equilibrium character of photoinduced superconductivity.   This more comprehensive picture could be obtained by performing frequency and angle resolved measurements.  However, these experiments would require more advanced instrumentation, including THz and IR free electron laser radiation.  Importantly, this would allow for the separate measurement of both signal and idler amplification.  This would provide direct access to the underlying non-linear dynamics of the photoinduced superconducting order parameter.

\section{Outlook}
\label{sec:outlook}

We envision several potentially interesting applications of the Higgs amplification phenomenon. Of particular interest for quantum information is the possibility of generation of entangled photon pairs at THz frequencies, as expected from Eq.~(\ref{Higgs_to_photons_schematic}).  Properties of the entangled photons may be controlled by tuning the intensity, duration, and angle of incidence of the pump beam. 

The notions introduced above can be generalized to the non-linear dynamics of other kinds of non-equilibrium condensates, including charge and spin density waves, and excitonic condensates.

Systems with several competing orders should exhibit multiple finite energy collective modes, leading to an additional richness of the order parameter dynamics.  The interaction between light and strongly excited collective modes opens a new frontier in the study of light-matter interaction  in many-body quantum states.

\acknowledgments

We thank R. Averitt, I. Carusotto, J. Faist,  M. Hatridge, A. Imamoglu, A. Georges, A. Millis, D. Pekker, and P. Zoller for illuminating discussions.  We acknowledge financial support from the European Research Council under the European Union's Seventh Framework Programme (FP7/2007-2013)/ERC Grant Agreement No. 319286 (QMAC), the Deutsche Forschungsgemeinschaft via the excellence cluster `The Hamburg Centre for Ultrafast Imaging—Structure, Dynamics and Control of Matter at the Atomic Scale' and the priority program SFB925, the National Science Foundation (NSF),  the Israel Science Foundation (grant 1803/18), the Harvard-MIT Center for Ultracold Atoms, DARPA DRINQS program (award D18AC00014), and the  Vannevar Bush Faculty Fellowship.  YW is supported by the Postdoctoral Fellowship of the Harvard-MPQ Center for Quantum Optics and the AFOSR-MURI Photonic Quantum Matter (award FA95501610323).  DP is grateful for the hospitality of ITAMP at
the Harvard-Smithsonian Center for Astrophysics, and of the Aspen Center for Physics,
which is supported by NSF grant PHY-1607611. 

\appendix

\section{Determining the optical conductivity of an equivalent homogenous medium}
\label{sec:homogeneous}

In the time-resolved experiments one measures the pump-induced difference in the complex reflected electric field $\Delta\widetilde E_r(\omega)$. The ``raw'' complex reflection coefficient in the photo-excited state $\widetilde r_\textrm{pumped}(\omega)$ can then be extracted by inverting the following equation:
\begin{equation}
	\frac{\Delta\widetilde E_r(\omega)}{\widetilde E_r^{0}(\omega)} = \frac{\widetilde r_\textrm{pumped}(\omega) - \widetilde r_0(\omega)}{\widetilde r_0(\omega)}
	\label{eq:dE}
\end{equation}
where $\widetilde r_0(\omega)$ is the unperturbed complex reflection coefficient of K$_3$C$_{60}$ known from broadband FTIR measurement and $\widetilde E_r^{0}(\omega)$ is the reflected electric field in the unperturbed state. If the pump light penetrates in the sample several times more than the probe light, one can assume that the probe pulse samples a volume in the material that has been homogenously transformed by the pump. In this case it is possible to directly extract the complex valued optical response functions by inverting the Fresnel equations.
 
\begin{figure*}[htbp]
    \includegraphics[width=0.9\textwidth]{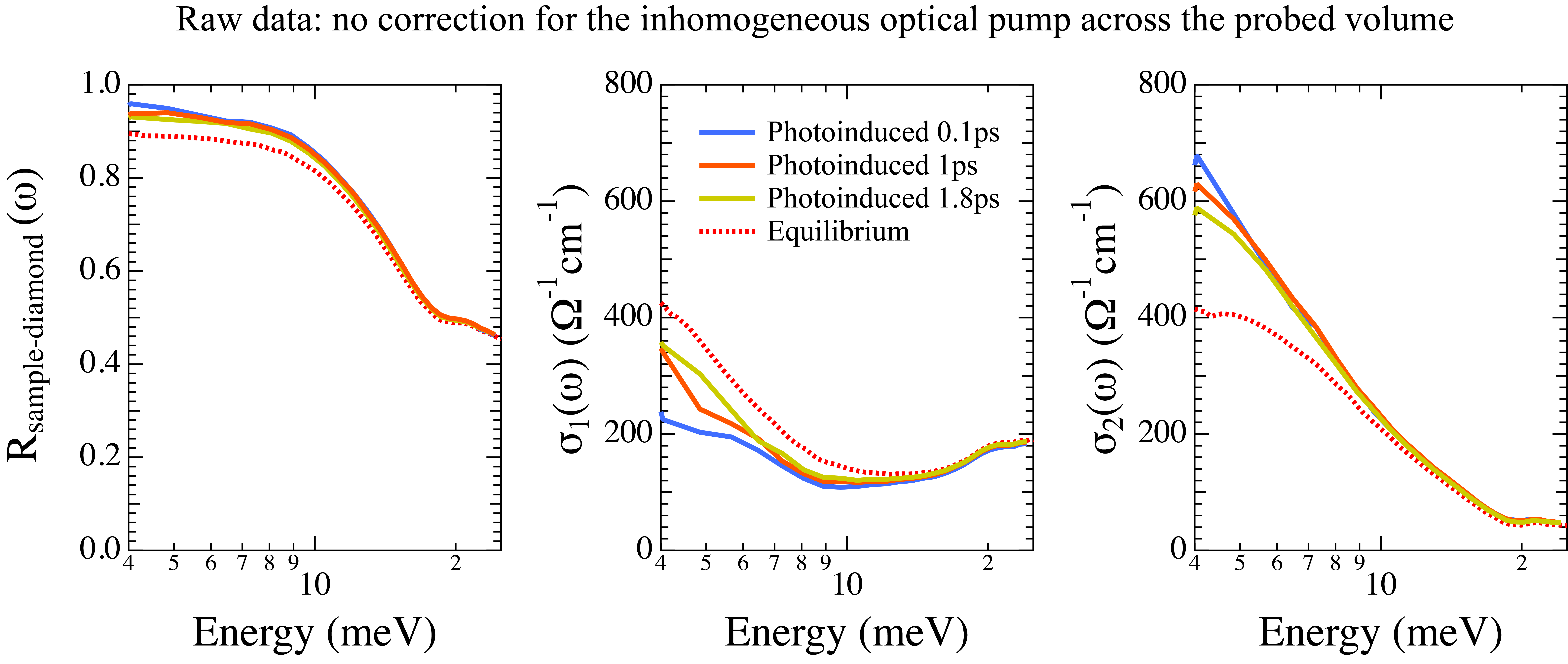}
    \caption{The left panel shows raw data for the reflectivity (sample-diamond interface), in equilibrium and after photoexcitation by pulses of three different lengths.  The second and third panels show the values of $\sigma_1(\omega)$ and $\sigma_2(\omega)$ that one would obtain from the observed reflectivity via the Fresnel equations, assuming optical conductivities independent of distance from the surface.  Note that this neglects that the intensity of the pump pulse decays as a function of distance from the surface of the sample and the probe is sampling an inhomogenously transformed volume.
        \label{fig:Raw}}
\end{figure*}

The conditions assumed in the previous paragraph are not correct for the experiments presented here, because the penetration depth of the mid-infrared pump (220 nm) is at least three times shorter than that of the THz probe (600-900 nm). Nevertheless, it is instructive to see what results for the optical conductivity would be obtained if the assumption is made.

In the left panel of Fig.~\ref{fig:Raw}, we show raw data for the reflectivity, in equilibrium and after photoexcitation by pulses of three different lengths.  In the second and third panels, we show the values of $\sigma_1(\omega)$ and $\sigma_2(\omega)$ that one would obtain from the observed reflectivities using the Fresnel equations, assuming optical conductivities independent of distance from the surface.

In contrast, the curves in Fig.~\ref{fig:ExpFig2} were obtained after taking account that the penetration depth of the probe radiation is longer than the that of the pump pulse, using the following approach.  As the pump penetrates in the material, its intensity is reduced and it will induce progressively weaker changes in the refractive index of the sample. This is modeled by ``slicing'' the probed thickness of the material into thin layers where we assume that the pump-induced changes in the refractive index are proportional to the pump intensity in the layer, {\em i.e.} $\widetilde n(\omega,z) = \widetilde n_0(\omega) + \Delta\widetilde n(\omega)e^{-\alpha z}$ where $n_0(\omega)$ is the unperturbed complex refractive index, $\alpha$ is the attenuation coefficient at the pump frequency, and $z$ is the spatial coordinate along the sample thickness.

For each probe frequency $\omega_i$, the complex reflection coefficient $\widetilde r(\Delta\widetilde n)$ of such multilayer stack is calculated with a characteristic matrix approach\cite{bornwolf} keeping $\Delta\widetilde n$ as a free parameter. As equation (\ref{eq:dE}) relates directly the measured quantity $\Delta\widetilde E_r(\omega)/\widetilde E_r^{0}(\omega)$ to the changes in reflectivity we can extract $\Delta\widetilde n$ by minimizing numerically:
\begin{equation}
	\left|\frac{\Delta\widetilde E_r(\omega_i)}{\widetilde E_r^{0}(\omega_i)} - \frac{\widetilde r(\omega_i,\Delta\widetilde n) - \widetilde r_0(\omega_i)}{\widetilde r_0(\omega_i)}\right|
\end{equation}
Note that $\Delta\widetilde n(\omega)$ represents the pump-induced change in the refractive index at the surface, where the pump has not yet been attenuated by the absorption in the material. By taking $\widetilde n(\omega) = \widetilde n_0(\omega) + \Delta\widetilde n(\omega)$ one can reconstruct the optical response functions of the material as if it had been homogeneously transformed by the pump.

The blue data points in Fig.~\ref{fig:ExpFig2} show the data processed in this manner to obtain the optical properties applicable to the region closest to the surface, where the excitation pulse is strongest.  Specifically, panels in the second and third columns show the effective values of $\sigma_1(\omega)$ and $\sigma_2(\omega)$ deduced for this region, while the first column shows the reflectivity that would be expected if these values of the optical conductivity were to hold independent of depth.  The blue solid curves show values obtained from the theoretical model with Higgs modulation.  Comparing Figs.~\ref{fig:ExpFig2} and \ref{fig:Raw}, we see that while the curves differ in detail,
the enhanced optical response for the shortest pulse data, seen in both figures, tends to support our theoretical model. In particular, while we do not actually observe amplification in the raw reflected signal, our analysis suggests that amplification would have been observed if the pump pulse were able to penetrate more deeply into the sample.

\section{Expressing the reflectivity in terms of a complex conductivity function}
\label{sec:RtoSig}
	
In the theory, we computed the complex reflection amplitude $r$, using Eq.~(\ref{eq:R11}).  To convert this to a conductivity function, we used the Fresnel relation for the reflection amplitude at the interface between two media
\begin{eqnarray}
r(\omega)=\frac{\sqrt{\epsilon_{\rm out}}-i\sqrt{-\epsilon(\omega)}}{\sqrt{\epsilon_{\rm out}}+i\sqrt{-\epsilon(\omega)}}\,.
\end{eqnarray}
For us, the first medium was diamond, which to a good approximation has a frequency-independent dielectric function $\sqrt{\epsilon_{\rm out}}=2.37$.  This relation was inverted,
\begin{eqnarray}
\epsilon(\omega)=\epsilon_{\rm out}\left(\frac{1-r(\omega)}{1+r(\omega)}\right)^2\,, \label{eq:r2sigma}
\end{eqnarray}
in order to obtain the complex conductivity using $\sigma(\omega)=-i\varepsilon_0\omega [\epsilon(\omega)-\epsilon_{\rm s}]$.

\section{Optical conductivity in the normal state}
\label{sec:sigmaEq}

\begin{figure}[]
	\centering
	\hspace{2pt}
	\begin{overpic}[width = 0.42\textwidth]{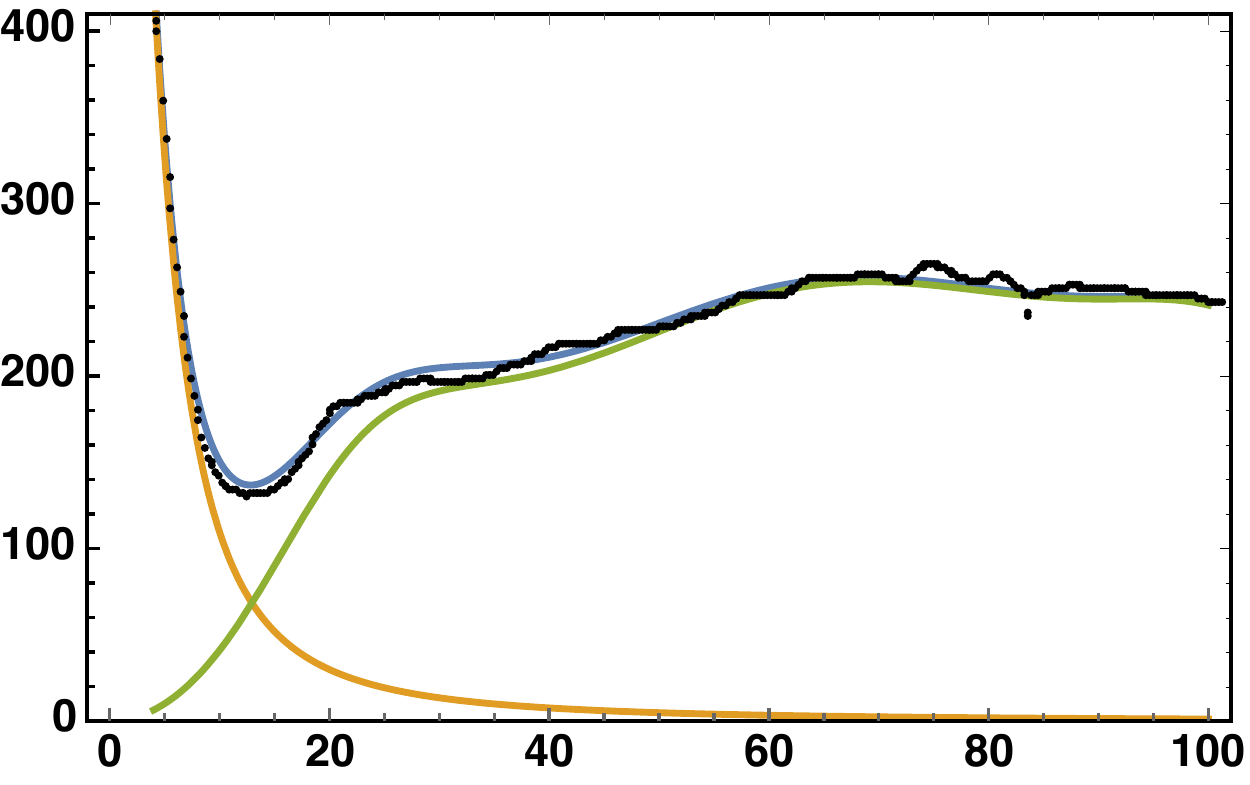}
			\put(-65,200){\rotatebox{90}{$\sigma_1$ $\left(\Omega^{-1}\mathrm{cm}^{-1}\right)$}}
			\put(770,-35){ $\omega$ (\mevh)}
			\put(310,260){\color{green} mid-infrared absorption peak}
			\put(305,90){\color{orange} Drude}
		\put(400,-60){(a) Real part}
	\end{overpic}
	\\
	\vspace{22pt}
	\hspace{2pt}	
	\begin{overpic}[width = 0.42\textwidth]{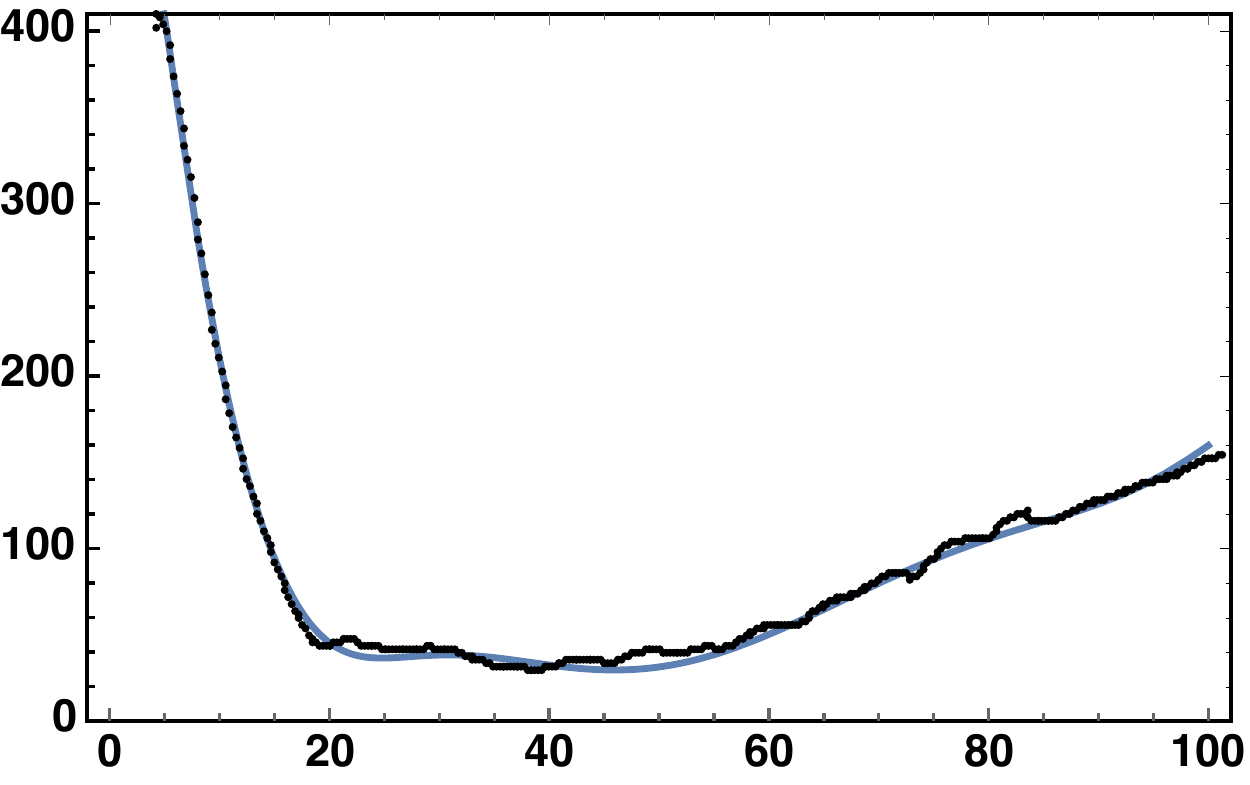}
			\put(-65,200){\rotatebox{90}{$\sigma_2$ $\left(\Omega^{-1}\mathrm{cm}^{-1}\right)$}}
			\put(770,-35){ $\omega$ (\mevh)}
			\put(370,-60){(b) Imaginary part}
	\end{overpic}
	\\
	\vspace{15pt}
	\caption{Optical conductivity of the normal state at equilibrium.   Black dots -- measured conductivity. Blue curve -- Fit to a sum of Drude and Lorentzian peaks, Eq.~(\ref{eq:sigmaEq}). Orange and green curves --  contributions to $\sigma_1$ arising from the Drude peak and Lorentzian peaks, respectively.  The Lorentzian peaks represent a broad mid-infrared absorption peak, which we assume for simplicity  not to be  affected by the onset of superconductivity.}
        \label{fig:normalFit} 
\end{figure}
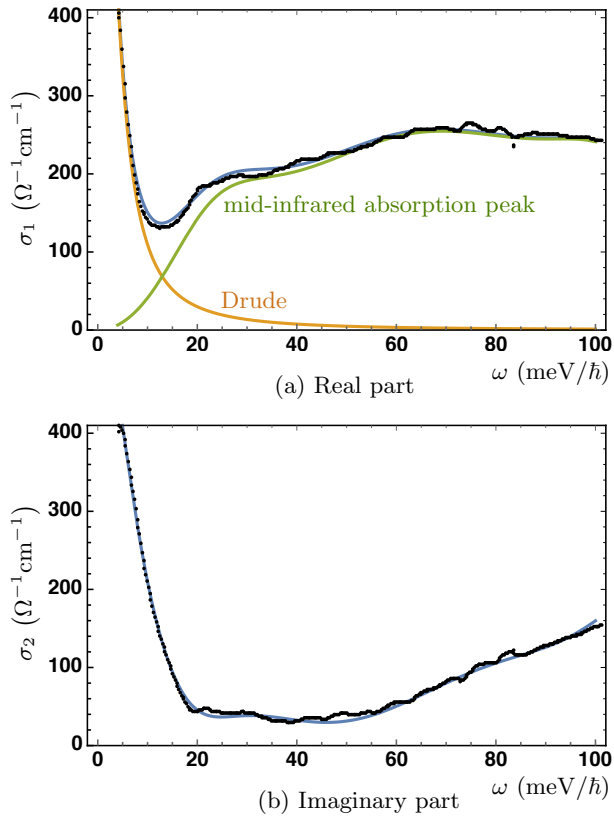

We model the optical conductivity $\tilde{\sigma}(\omega)$ of the equilibrium normal state as the sum of a Drude peak, representing the conduction band, and a sum of three Lorentzians,  representing a broad mid-infrared absorption peak \cite{RiceChoi}.
\begin{eqnarray}
\sigma_{\rm eq}(\omega)=\frac{\Lambda_{\rm s,eq}}{\gamma_{\rm D}-i \omega} + \sum_{n=1}^3  \frac{B_n\, \omega}{i(\Omega_n^2-\omega^2)+ \gamma_n\, \omega} \label{eq:sigmaEq}
\end{eqnarray}
A fit to the optical conductivity of the equilibrium normal state in the measured range $4 \mevh<\omega<100 \mevh$ (see Figure \ref{fig:normalFit}) gives the following values: $\Lambda_{\rm s,eq}=3,470\, \Omega^{-1}\mathrm{cm}^{-1}\mevh$, $\gamma_{\rm D}=3.56\,\mevh$, $B_1=18,300\,\Omega^{-1}\mathrm{cm}^{-1}\mevh$, $\Omega_1=70.4\,\mevh$, $\gamma_1=86.6\,\mevh$,  $B_2=4,600\,\Omega^{-1}\mathrm{cm}^{-1}\mevh$, $\Omega_2=26.1\,\mevh$, $\gamma_2=34.0\,\mevh$,  $B_3=2,400\,\Omega^{-1}\mathrm{cm}^{-1}\mevh$, $\Omega_3=102.6\,\mevh$, $\gamma_3=35.0\,\mevh$.

\end{document}